\documentstyle[pre,epsfig,eqsecnum,aps]{revtex}
\begin{document}
\title{Critical light scattering in liquids.}

\author{G. Flossmann and R.~Folk}

\address{Institute for Theoretical Physics, University of Linz, Austria}

\maketitle

\begin{abstract}
We compare theoretical results for the characteristic frequency of the
Rayleigh peak calculated in one-loop order within the field theoretical
method of the renormalization group theory with experiments and other
theoretical results. Our expressions describe the non-asymptotic
crossover in temperature, density and wave vector. In addition we
discuss the frequency dependent shear viscosity evaluated within the
same model and compare our theoretical results with recent
experiments in microgravity.
\end{abstract}

\vspace{1cm}
\centerline{PACS: 64.60.Ht, 05.70.Jk, 62.60.+v, 64.70.Fx}
\today

\section{Introduction}

Dynamical critical phenomena manifest themselves in a singular temperature
dependence of hydrodynamic transport coefficients \cite{HH77}. In pure
fluids these transport coefficients are the thermal conductivity and the shear
viscosity, both diverging on approach of the critical point. In Ref.~\cite{fomo}
the field theoretic renormalization group (RG) theory has been used for a quantitative
description of this non-analytic behavior and attention was given to the crossover to
the analytic behavior in the background further away from $T_c$. The thermal conductivity
can be measured in two ways, (i) by measuring the temperature difference when a heat
current flows through the liquid (this is an experiment at zero wave vector $k$), and (ii) by
light scattering experiments in the hydrodynamic region, where the wave vector and the
temperature dependent correlation length $\xi$ fullfil the relation $k\xi<<1$. In the last
case the thermal diffusivity $D_T$ is measured which is related to the thermal conductivity
$\kappa_T$ by $D_T={\kappa_T}/{(\rho C_P)}$ so that for a comparison of the two experimental
data the specific heat has to be known. For the thermal diffusivity and the shear viscosity
however theoretical calculations show that no other static quantity apart from the
correlation length has to be known. This makes these two transport coefficients most suitable
to check the dynamical renormalization calculation. 

In light scattering experiments in liquids the characteristic frequency $\omega_c$, defined as the
half width at half height of the central Rayleigh peak, provides useful
additional information  about the dynamical properties of the system.
Far away from the critical point in the hydrodynamic region the characteristic
frequency is given by $\omega_c=D_T(T,\rho)\,k^2$. Approaching the
critical point a crossover from the hydrodynamic to the so called critical region
($k\xi>>1$) takes place and finally at ($T_c,\rho_c$) the characteristic frequency is 
a function of the wave vector alone. Asymptotically near the critical
point (for very small values of $k$) the power law behavior $\omega_c\sim k^z$ is expected
with the dynamical critical exponent $z\approx 3$. Further away, that means for larger wave vector
modulus, a crossover to the background behavior with a non singular thermal
conductivity takes place. This is described by the van Hove theory where 
the characteristic frequency behaves as $\omega_c\sim k^4$.

It is the aim of this paper to calculate the characteristic frequency
in the whole ($\xi,k)$-plane within the non-asymptotic RG
theory in order to describe all types of crossover quantitatively.
In addition the density dependence of the line width is considered.
The non-universal background parameters entering the expression for the
characteristic frequency are taken from other dynamical experiments, e.g
measurements of the shear viscosity. Recently very precise data became
available for xenon from experiments performed in microgravity \cite{BMZ}.
This allows also to reconsider the frequency dependence of the shear
viscosity within RG theory already discussed in Ref.~\cite{fomoflo}.

The results for pure fluids are also compared with light scattering experiments in 
polymer solutions and polymer blends. The non-asymptotic behavior in a mixture is not
completely described by the critical model for pure fluids \cite{fomo2} but
the asymptotics is the same. Therefore agreement should be found as long as
the non-universal dynamic parameters are near to their fixed point values.

\section{The dynamic model}

The dynamic order parameter correlation function for the gas-liquid transition can be described
within the model H \cite{HH77}, which is a special case of the model H' described in detail in Ref.~\cite{fomo},
containing dynamic equations for the order parameter $\phi_0$ (the entropy density)
and the transverse momentum density $\mbox{\boldmath$j$}_t$, 
\begin{eqnarray}
\frac{\partial\phi_0}{\partial t} &=&
\stackrel{o}{\Gamma}\nabla^2\frac{\delta H}{\delta\phi_0}
-\stackrel{o}{g}(\mbox{\boldmath$\nabla$}\phi_0)\ \frac{\delta H}
{\delta\mbox{\boldmath$j$}_l} +\Theta_\phi \,,\label{dphidt}\\
\frac{\partial\mbox{\boldmath$j$}_t}{\partial t}&=&
\stackrel{o}{\lambda}_t\nabla^2\frac{\delta H}{\delta\mbox{\boldmath$j$}_t}+
\stackrel{o}{g}{\cal T}\left\{
(\mbox{\boldmath$\nabla$}\phi_0)\frac{\delta H}{\delta\phi_0}
-\sum_{k}\left[j_k\mbox{\boldmath$\nabla$}
\frac{\delta H}{\delta j_k}-\nabla_k\mbox{\boldmath$j$}
\frac{\delta H}{\delta j_k}\right]\right\}+\mbox{\boldmath$\Theta$}_t \,,\label{djtdt}
\end{eqnarray}
with fast fluctuating forces $\Theta_i$ and the projector ${\cal T}$ to the direction of the transverse
momentum density. The Hamiltonian appearing in the dynamic equations is the normal Hamiltonian
of a $\phi^4$-theory together with the conserved density $\mbox{\boldmath$j$}_t$ entering
quadratically:
\begin{eqnarray}\label{hamil}
H&=&\int d^dx\left\{\frac{1}{2}\stackrel{o}{\tau}\phi_0^2
+\frac{1}{2}(\mbox{\boldmath$\nabla$}\phi_0)^2+
\frac{\stackrel{o}{\tilde{u}}}{4!}\phi_0^4+\frac{1}{2}a_j\mbox{\boldmath$j$}^2_t\right\}\ .
\end{eqnarray}
As described in Ref.~\cite{fomo} the dynamic equations may be transformed into a dynamic functional leading
to dynamic vertex functions which can be calculated in perturbation theory.
In general the dynamic scattering function is different from a Lorentzian
due to fluctuation effects. This prediction of scaling theory has been
observed in ferromagnets \cite{ferro} and even compared with RG
calculations \cite{iro}. The same scaling arguments as for the ferromagnet also
apply for pure fluids although the deviation from a Lorentzian is expected to be
smaller \cite{sofia}. Moreover it turns out that in one-loop order there
are no frequency dependent contributions in the one-loop perturbation
terms of the order parameter vertex functions \cite{remark1}. Therefore 
the shape of the dynamic correlation function is approximated by a
Lorentzian and may be written as
\begin{equation}
\chi_{dyn}(k,\xi,\omega) = 2\chi_{st}\Re\left[\stackrel{o}{\Gamma}_{\phi\tilde\phi}^{-1}(k,\xi,\omega)\right]
= \frac{\chi_{st}(k,\xi)}{\omega_c(k,\xi)}\,
\frac{2}{1+y^2}\label{chidyn}
\end{equation}
in one-loop order with $y=\omega/\omega_c$ and the characteristic frequency $\omega_c$, defined
as the half width at half height of the Rayleigh peak. The width is given by the vertex
function $\Gamma_{\phi\tilde\phi}(k,\xi,\omega\!=\!0)$ so that the unrenormalized
characteristic frequency reads
\begin{equation}
\stackrel{o}{\omega}_{c}(k,\xi) = \,\,\stackrel{\circ}{\Gamma}\!k^2(\xi^{-2}+k^2)
\left(1+\frac{\stackrel{\circ}{f}_t^2}{\xi^{d-4}} \int d^dp\,\frac{1}
{1+({\bf x}-{\bf p})^2}\,\frac{\sin^2\theta}{p^2}\right) \,,\label{vertex2a} 
\end{equation}
with $x=k\xi$, $\Omega=\omega/\!\!\stackrel{\circ}{\Gamma}$ and $\stackrel{\circ}{f}_t=\,\,\stackrel{\circ}{g}\!\!/
\sqrt{\stackrel{\circ}{\Gamma}\stackrel{\circ}{\lambda}_t}$ after setting the parameter
$\stackrel{\circ}{w}_\phi=\,\,\stackrel{\circ}{\Gamma}\!\!/a_j\!\!\stackrel{\circ}{\lambda}_t$, which is irrelevant
under renormalization, to zero. In full analogy to the renormalization of the transport coefficients
\cite{fomo} the pole in the unrenormalized characteristic frequency may be absorbed into Z-factors using
field theoretic renormalization group theory. As we get the same Z-factors (and thus the same flow
equations for the Onsager coefficient and the mode coupling) as for the transport
coefficients we shall skip the details here.

\section{The characteristic frequency}

\subsection{General expression}
After renormalization the characteristic frequency $\omega_c$ is finally found to be
\begin{eqnarray}
\omega_c(k,\xi) = \Gamma(\ell)\,k^2(\xi^{-2}+k^2)
\left\{1 - \frac{f_t^2(\ell)}{16}\,\left[-5 + 6\,x^{-2}\,\ln(1+x^2)\right]\right\}  \label{omegachar0}
\end{eqnarray}
in an $\epsilon$-expansion with $\epsilon=4-d$. The temperature dependence enters via the flow equations for the mode coupling
and the Onsager coefficient,
\begin{eqnarray}
f_t^2(\ell) &=& f_t^{*2}\left[1+\frac{\ell}{\ell_0}\left(\frac{f_t^{*2}}{f_0^2}-1\right)\right]^{-1} \,,
\label{modecoupling} \\
\Gamma(\ell) &=& \Gamma_0\left(\frac{f_0^2}{f_t^{*2}}\frac{\ell_0}{\ell}
\left[1+\frac{\ell}{\ell_0}\left(\frac{f_t^{*2}}{f_0^2}-1\right)\right]\right)^{1-x_\eta} \label{onsager} \,,
\end{eqnarray}
with the one-loop fixed point value of the mode coupling
$f_t^{*2}=\frac{24}{19}$ and the one-loop value of the exponent $x_\eta=\frac{1}{19}$.
The connection between the flow
parameter $\ell$ and the correlation length or the wave vector
respectively is found from the matching condition
\begin{equation}
\left(\xi_0^{-1}\ell\right)^2 = \xi^{-2}+k^2 \,,
\label{matchgeneral}
\end{equation}
for the Lorentzian approximation where the correlation length may be expressed in terms of the reduced temperature
$t$ via $\xi=\xi_0\,t^{\nu}$ with $\nu=0.63$ along the critical isochore. As described in Ref.~\cite{fomoflo} we may use
the cubic model to include non-critical values of the reduced density. In (\ref{modecoupling})-(\ref{matchgeneral})
$\Gamma_0$ and $f_0$ are the initial values of the Onsager coefficient and the mode coupling at an arbitrary reduced
temperature $t_0$ along the critical isochore, $\ell_0$ is the solution of the matching condition at $t_0$ and $k=0$
and $\xi_0$ is the amplitude of the correlation length. Eq. (\ref{matchgeneral}) is the frequency independent matching
condition which has been used since the vertex function $\Gamma_{\phi\tilde\phi}$, expressing the characteristic
frequency in the Lorentzian approximation, is evaluated at zero frequency.

We may rewrite (\ref{omegachar0}) extracting the asymptotic expressions for the Onsager coefficient
and the mode coupling, 
\begin{equation}
\omega_c(k,x) = \Gamma_{as}\,k^z\left(\frac{1+x^2}{x^2}\right)^{1-x_\lambda/2}\!\!\!\!
\left[c_{na}(k,x)\right]^{x_\lambda}f(k,x)\label{omegachar} 
\end{equation}
with $x=k\xi$, $z=4-x_\lambda$ and the function $f(k,x)$ defined as
\begin{equation} 
f(k,x)= 1 - \frac{f_t^{*2}}{16 \,  c_{na}(k,x)}\,\left[-5 + 6\,x^{-2}\,\ln(1+x^2)\right]\,.
\end{equation}
The non-asymptotic contributions are collected in
\begin{equation} 
c_{na}(k,x)=\left[1+\frac{k}{k_0}\,\sqrt{\frac{1+x^2}{x^2}}\,\right]  
\end{equation}
so that the asymptotic region is characterized by $c_{na}(k,x)=1$. Finally the asymptotic
Onsager coefficient $\Gamma_{as}$ and the crossover wave length $k_0$ are given by
\begin{equation} \label{gammaas}
\Gamma_{as} = \Gamma_{0}\left(
\frac{f_0^2\ell_0}{f_t^{*2}\xi_0}\right)^{x_\lambda}, \qquad
k_0^{-1}=\left(\frac{f_t^{*2}}{f_0^2}-1\right)\frac{\xi_0}{\ell_0} \,\, .
\end{equation}  
The advantage of Eq. (\ref{omegachar}) over Eq. (\ref{omegachar0}) is the clear separation of the asymptotic
and the non-asymptotic behavior which will make the discussion of the various limits of the characteristic
frequency easier. Before we come to that point in the next section we should remark here that it is also possible to
evaluate the crossover function in three dimensions \cite{SD} instead of performing an $\epsilon$-expansion.
The characteristic frequency then reads
\begin{eqnarray}
\omega_c(k,\xi) = \Gamma(\ell)\,k^2(\xi^{-2}+k^2)
\left\{1 + f_t^2(\ell)\left[\frac{2}{3}\,\sqrt{\frac{1+x^2}{x^2}}\,\arctan x - \frac{3}{4}\right]\right\}
\label{omegachar1}
\end{eqnarray}
As expressions (\ref{omegachar0}) and (\ref{omegachar1}) are almost identical after choosing the right
initial values for the Onsager coefficient and the mode coupling we shall only discuss the $\epsilon$-expansion
result (also used for the evaluation of the transport coefficients in \cite{fomo,fomoflo}) in the following.

\subsection{Various limits of the characteristic frequency}

First we should note that Eq. (\ref{omegachar}) yields a finite value for
the characteristic frequency in the {\it hydrodynamic} limit $x\to0$,
\begin{equation}
\lim_{x\to0} \omega_c(k,\xi) = \Gamma_{as}\,k^2\,\xi^{-2+x_\lambda}\!
\left[1+\frac{1}{x_0}\right]^{x_\lambda}\left\{1 - \frac{f_t^{*2}}{16}\,
\left[1+\frac{1}{x_0}\right]^{-1}\right\}\,, \label{hydrolimit}
\end{equation}
with $x_0=k_0\xi$. Here the coefficient of $k^2$ is the non-asymptotic expression for the temperature
dependent thermal diffusion coefficient $D_T(\xi)$ discussed in Ref.~\cite{fomo,fomoflo} so that we can
rewrite Eq.~\ref{hydrolimit} in the well-known form $\omega_c=D_T(\xi)k^2$ for the hydrodynamic region.
Also in the opposite {\it critical} limit $x\to\infty$ we obtain a finite value for the characteristic
frequency,
\begin{equation} 
\lim_{x\to\infty} \omega_c(k,\xi) =\omega_c(k)=
\Gamma_{as}\,k^z\,\left[1+\frac{k}{k_0}\right]^{x_\lambda}
\left\{1 + \frac{5f_t^{*2}}{16}\,\left[1+\frac{k}{k_0}\right]^{-1}\right\}\,,
\end{equation}
which is the wave vector dependent non-asymptotic expression of the characteristic frequency. 
Both non-asymptotic expressions allow to discuss the crossover from the {\it asymptotic} limit
$\xi k_0\to \infty$ or $k/k_0\to 0$ to the {\it background} limit $\xi k_0\to 0$ or $k/k_0\to \infty$
respectively.

In the hydrodynamic case we obtain the limits
\begin{eqnarray}
\lim\limits_{\xi k_0\to \infty}\omega_c(k,\xi) &=&
\Gamma_{as}\,k^2\,\xi^{-2+x_\lambda}\left(1-\frac{f_t^{*2}}{16}\right) \,,\label{limit1}\\
\lim\limits_{\xi k_0\to 0}\omega_c(k,\xi) &=&
\Gamma_{0}\,k^2\,\xi^{-2}\left(1-\frac{f_0^2}{f_t^{*2}}\right)^{x_\lambda} \,, \label{limit2}
\end{eqnarray}
where we used the expression for $k_0$ given in Eq.(\ref{gammaas}) for the last limit. In the background
limit our expression reaches the van Hove behavior. In the  critical region we obtain
\begin{eqnarray} 
\lim\limits_{k/k_0\to0}\omega_{c}(k) &=&
\Gamma_{as}\,k^{z}\left(1+\frac{5 f_t^{*2}}{16}\right)\,, \label{limit3}\\
\lim\limits_{k/k_0\to\infty}\omega_{c}(k) &=&
\Gamma_{0}\,k^{4}\left(1-\frac{f_0^2}{f_t^{*2}}\right)^{x_\lambda}\,, \label{limit4}
\end{eqnarray}
where again we reach the van Hove theory for large values of the ratio $k/k_0$. This means that our
results for the characteristic frequency describe the crossover in the correlation length
(from $\xi^{-2+x_\lambda}$ to $\xi^{-2}$) in the hydrodynamic region characterized by the limit $x\to0$
as well as the crossover in the wave vector (from $k^z$ to $k^4$) in the critical region characterized
by the limit $x\to\infty$.

We have seen that with our non-asymptotic theory we always reach the van Hove behavior in the
non-asymptotic limit for large values of the wave vector or small values or the correlation length
respectively. This is different from the non-asymptotic mode coupling expression of
Olchowy \cite{LSO95}, where the characteristic frequency is given by
\begin{equation}
\omega_c(k,\xi) = \frac{k_BT}{6\pi\bar\eta^B\xi}\,k^2\,
\frac{3}{4}\,(1+x^2)^{-1/2}\,\left[-y_D+y_\delta(1+x^2)^{1/2}\right]\label{firstolch}
\end{equation}
with
\begin{equation}
y_D=\arctan x_D\,,\qquad
y_\delta = (1+x_D^2)^{-1/2}\,\left[-y_D+\arctan(x_D(1+x_D^2)^{-1/2}\right]
\end{equation}
depending both on the non-universal parameter $x_D=q_D\xi$ which is similar to the parameter $k_0$
appearing in our non-asymptotic theory. Eq. (\ref{firstolch}) does not yield the van Hove theory
in the non-asymptotic region but instead becomes negative for $x > 2 x_D$. This region of unphysical
negative values of the characteristic frequency is always reached at constant correlation length when
the wave vector becomes larger than the non-universal parameter $q_D$. On the other hand the parameter $q_D$
cannot be set to infinity as this limit yields an unphysical divergence in the hydrodynamic limit
for $\xi\to0$ \cite{LSO95}.

\subsection{Discussion of the crossover behavior}

In the background we always reach the van Hove behavior for the characteristic frequency.
This is a general feature of our non-asymptotic theory. The parameter which describes the
crossover from the van Hove expression of the characteristic frequency to its asymptotic expression
is in fact the value of the mode coupling $f_0$ which can take on values from zero to the fixed point
value $f_t^{*}$. Note that this corresponds to a crossover of $k_0$ from its asymptotic limit $k_0\to \infty$
to its van Hove limit $k_0\to 0$. The van Hove behavior for $f_0=0$,
\begin{equation} \label{vanhove}
\omega^{vH}_{c}(k,x)=\Gamma_{0}\,k^{4}\left(1+x^{-2}\right) \, ,
\end{equation} 
is different from the background behavior at finite $f_0$ so that we can define a background van Hove characteristic
frequency $\omega^{BvH}_c$ as
\begin{equation}
\omega^{BvH}_{c}(k,x)=\Gamma_{0}\,k^{4}\left(1-\frac{f_0^2}{f_t^{*2}}\right)^{x_\lambda}\left(1+x^{-2}\right) \, ,
\end{equation}
which we now always reach with our non-asymptotic theory in the background limit $\xi k_0\to0$ or $k/k_0\to\infty$
respectively.

Now we can extract the background van Hove behavior from the full characteristic frequency given in Eq.~\ref{omegachar},
\begin{equation}
\omega_c(k,x)=\omega^{BvH}_{c}(k,x)\left(k\xi_0\right)^{-x_\lambda}
\left(\frac{f_0^2}{f_t^{*2}}\,
\ell_0\right)^{x_\lambda}\left(\frac{1+x^2}{x^2}\right)^{-x_\lambda/2}
\left(\frac{c_{na}(k,x)}{1-\frac{f_0^2}{f_t^{*2}}}\right)^{x_\lambda}f(k,x), 
\end{equation}
and plot the ratio $\omega_c/\omega_c^{BvH}$ in order to demonstrate the crossover behavior of the
characteristic line width. This is done in Fig.~\ref{flossf3a} from which we see that the ratio
$\omega_c/\omega_c^{BvH}$ increases near the critical point (characterized by $\xi\to\infty$ and
$k\to0$) as the characteristic frequency then approaches its asymptotic power law behavior, of
course with nonuniversal amplitudes depending on value of the mode coupling $f_0$ in the background.
This effect increases with increasing values of the mode coupling $f_0$. Especially we see that choosing
the fixed point value $f_0=f_t^{*}$ the surface of the characteristic frequency never reaches a flat surface
(corresponding to the van-Hove behavior).

The crossover from the asymptotic power-law behavior in the critical region, where the
characteristic frequency is proportional to $k^z$,  to the van Hove behavior with $\omega_c\propto k^4$
in the non-asymptotic background region can also be seen in Fig.~\ref{flossf03} where we compare
our asymptotic and non-asymptotic results with the van Hove theory and the result of
Kawasaki \cite{KL72}. To do this we rewrite Eq. (\ref{omegachar}) extracting $k^2$ instead of $k^z$,
\begin{equation}
\omega_c^{as}(k,\xi) = \frac{\Gamma_{as}}{\xi^{1+x_\eta}}\,k^2 (1+x^2)^{1-x_\lambda/2}
\left[c_{na}(k,x)\right]^{x_\lambda}f(k,x)
\equiv \frac{\Gamma_{as}}{\xi^{1+x_\eta}}\,k^2\,\Omega(x) \,,\label{omegachar2}
\end{equation}
and compare the various results for the function $\Omega(x)/x$ at constant correlation length
instead of $\omega_c$ itself. Therefore we have to note that the function $\Omega(x)$ defined in
Eq. (\ref{omegachar2}) is in general not only a function of $x=k\xi$ but also of the
wave vector $k$ which enters via the non-asymptotic function $c_{na}(k,x)$. But keeping the
correlation length constant as in Fig.~\ref{flossf03} we can express $k$ in terms of $x$
so that $\Omega(x)$ is really only a function of $x$. 

As Kawasaki's result is proportional to $k^3$ instead of $k^z$ the function $\Omega(x)/x$ plotted
in Fig.~\ref{flossf03} becomes constant for large values of $x$ whereas our asymptotic result
(characterized by $c_{na}(k,x)=1$) is proportional to $x^{x_\eta}$ and the van Hove theory to $x$.
Our non-asymptotic results (at constant values of the correlation length $\xi$) behave for large values of $x$ like
the van Hove theory and are therefore proportional to $x$. We also see in this figure that
the setin of the crossover to the van Hove theory is determined by initial value of the mode coupling
$f_0$ which is the only free parameter in our non-asymptotic theory. We also should note that
in Kawasaki's theory there is a different prefactor for the function $\Omega(x)$ so that we have
normalized the function $\Omega(x)/x$ so that the curves coincide for $x\to0$. 

We can also use the function $\Omega(x)$ defined in Eq. (\ref{omegachar2}) to compare our asymptotic
result for the characteristic frequency with other theories: In Fig.~\ref{flossf04} we have plotted
the asymptotic result for $\Omega(x)/x$ (which is only a function of $x$ as we have $c_{na}(k,x)=1$)
as well as the theoretical results of Kawasaki and Lo \cite{KL72}, Paladin and Peliti \cite{PP82} and
Burstyn et al. \cite{BSBF83}. As the other authors have a different prefactor
for $\Omega(x)$ we have normalized $\Omega(x)$ so that the curves coincide for $x\to0$. Again we see that the Kawasaki
result for $\Omega(x)/x$ becomes constant whereas the other results show the correct $x^{x_\eta}$ behavior for large values
of $x$. In addition to this comparison we should note that at the critical dimension $d=4$ our result for the characteristic
frequency is identical with the result of Siggia et al. \cite{SHH76}.  

And finally let us mention that we can extend our theory to non-critical values of the density and
calculate the crossover in the characteristic frequency when we leave the critical isochore:
Using the parametric representation to connect the correlation length to the reduced
temperature $t=(T-T_c)/T_c$ and the reduced density $\Delta\rho=(\rho-\rho_c)/\rho_c$
\cite{fomoflo} we are able to evaluate the correlation length as a function of $t$,
$\Delta\rho$ and $k$. In Fig.~\ref{flossf02} we have plotted the characteristic frequency
in the hydrodynamic limit for $k=0$ as a function of $t$ and $\Delta\rho$. We see that the
characteristic frequency goes to zero in the critical limit $t\to0$ and $\Delta\rho\to0$
corresponding to $\xi\to\infty$.

\section{Comparison with experiments}

\subsection{Pure liquids}

In Fig.~\ref{flossf05}-\ref{flossf10} we compare our asymptotic and non-asymptotic results for the
characteristic frequency $\omega_c/k^2$ as a function of the reduced temperature $t$ and the function
$\Omega(x)/x$ as a function of $x$ with experiments in Xe and ${\rm CO_2}$ \cite{SH73} (all non-universal
parameters are given in Tab.~\ref{tab}). As discussed in Ref.~\cite{fomoflo} we can
treat the exponent $x_\lambda=1-x_\eta$ as an additional free parameter so that we can fit $f_0$ and $x_\eta$ from
the experimental data (the initial value of the Onsager coefficient $\Gamma_0$ is determined by the
value of the shear viscosity at $t_0$). But this means that we need additional data for this fit.
In Xe we have used the recent shear viscosity data of Berg et al. \cite{BMZ} discussed in the next section.
Fitting the parameter $f_0$ from the characteristic frequency data (the exact value of $x_\eta$ does hardly affect
the exponent $x_\lambda=1-x_\eta$) and the exponent $x_\eta$ from the shear viscosity data we find good agreement
for the characteristic frequency (Fig.~\ref{flossf05} and \ref{flossf06}) as well as for the frequency dependent
shear viscosity (Fig.~\ref{flossf07} and \ref{flossf08}). For ${\rm CO_2}$  we have taken $t_0$, $f_0$ and $\Gamma_0$
(also given in Tab.~\ref{tab}) from the comparison of the shear viscosity and the thermal diffusivity with experiments
in Ref.~\cite{fomoflo} so that the curves shown in Fig.~\ref{flossf09} and \ref{flossf10} are obtained without any free parameter!

As we can see in these figures the experimental data are not described correctly by our asymptotic expressions
but only by the non-asymptotic expressions which show the crossover to the van Hove theory for large values of
the reduced temperature $t$ or small values of the variable $x$ respectively. Analogously any asymptotic theory
\cite{KL72,PP82,BSBF83} fails to describe the experimental data correctly. In Ref.~\cite{SH73} this problem was eliminated
adding a regular background contribution of the form $\omega_c^B=(\lambda^B/\rho c_p)k^2(1+x^2)$ to the critical expression
for the characteristic frequency with $\lambda^B$ being the regular part of the thermal conductivity and $c_p$ the full
specific heat at constant pressure containing also critical contributions. The use of the full specific heat together with
the term $1+x^2$ ensures the crossover to the van Hove theory for large values of the reduced temperature as well as for
large values of the wave vector (the background characteristic frequency is proportional to $k^2\xi^{-2}$ for $x\to0$
and to $k^4$ for $x\to\infty$) so that the full characteristic frequency $\omega_c = \omega_c^C+\omega_c^B$ obtained
by this procedure yields basically the same curves as our non-asymptotic theory (see Fig.~6 of Ref.~\cite{SH73}). In our
theory however we use a different form of the background characteristic frequency: Following the discussion of the regular
background added to the transport coefficients \cite{fomoflo} we would have to add a background of the form
$\omega^B = D_T^B(T,\rho)k^2-D_T^B(T_c,\rho_c)k^2$ to our results with the background thermal diffusivity given by
$D_T^B = \lambda^B/\rho c_p^B$ and the background specific heat $c_p^B$ containing only the regular temperature
dependence without the critical singularity. As this background term turns out to be negligibly small in the
temperature range shown in Fig.~\ref{flossf05}-\ref{flossf10} we have neglected it so that our asymptotic and
non-asymptotic curves for Xe and ${\rm CO_2}$ contain only the critical contributions discussed in this paper.
So the main difference between our non-asymptotic theory and the results of Ref.~\cite{KL72,PP82,BSBF83} is that the
crossover to the van Hove theory, which is clearly seen in experiments, is already contained in our expressions
for the characteristic frequency and not added by an appropriate form of the background contribution!

In Fig.~\ref{flossf06} and \ref{flossf10} we also see that the non-asymptotic results for $\Omega(x)/x$ do
of course not collapse on a single curve (in contrary to our asymptotic result and the theories
of Ref.~\cite{KL72,PP82,BSBF83}) as the non-asymptotic contribution $c_{na}$ does not only depend on
the variable $x=k\xi$ but also on the wave vector $k$ and the correlation length $\xi$ separately.
This behavior can also be seen in the  Xe and ${\rm CO_2}$ data in Fig.~\ref{flossf06} and \ref{flossf10}
although the experimental data are not precise enough for a true confirmation of the validity of
our non-asymptotic theory.

\subsection{Polymer solutions and blends}

And finally we apply our theory for the characteristic frequency to light scattering experiments
in binary polymer solutions: In Fig.~\ref{flossf11} we compare our non-asymptotic theory for
the characteristic frequency $\omega_c/k^2$ as a function or the reduced temperature with
experimental data in a solution of polydisperse polystyrene (PDPS) in cyclohexane \cite{KKD98}.
For this figure the initial value of the Onsager coefficient was determined from the value
of the background shear viscosity at the critical point also measured in Ref.~\cite{KKD98}. The
amplitude of the correlation length $\xi_0$ as well as the exponents $\nu=0.7$ and $x_\eta=0.065$
were taken from the same paper. Therefore we have to note that the exponent $\nu$ found by
Ref.~\cite{KKD98} for the polymer solution is higher than the value $\nu=0.63$ found for pure liquids
or liquid mixtures. Fitting the initial value of the mode coupling $f_0$, which is the
only free parameter in our theory, from the experimental data (all values given in Tab.~\ref{tab})
we reach a satisfactory description of the experimental data although the curves for large wave vectors
lie above the experimental data for small values of the reduced temperature. Nevertheless we have
to note that the quality of the description cannot be compared to the one reached for
Xe and ${\rm CO_2}$ as there are no detailed experimental data for the shear viscosity
of this polymer solution in the vicinity of the critical point available so that an
exact determination of $\bar\eta_0$ and thus of $\Gamma_0$ was not possible and also
the critical exponent had to be fixed and could not be fitted from the experiments.

However one crucial point remains: The fact the we have used the non-asymptotic theory developed
for pure liquids to describe a polymer solution is of course a problem as liquids and
liquid mixtures do have the same asymptotic behavior but show a slightly different crossover to
the non-asymptotic behavior. But as an asymptotic theory is not able to describe the experimental
data (in the same way as we were not able to describe the characteristic frequency in pure liquids
with the asymptotic theory) and a non-asymptotic theory for critical light scattering in mixtures
has not yet been set up we believe that the systematic errors made by applying a non-asymptotic
theory for pure liquids to mixtures (which basically means setting the additional parameter $w_3$
found for the transport coefficients in liquid mixtures \cite{fomo2} to zero) are rather small
and can be tolerated. In addition we have to note that a background characteristic frequency given
in Ref.~\cite{KKD98} was subtracted from the experimental data as well as from the non-asymptotic results
for $\omega_c/k^2$.

In Fig.~\ref{flossf12} we compare our asymptotic result for the function $\Omega(x)/x$ as well
as the theoretical results of Kawasaki \cite{KL72} and Burstyn et al. \cite{BSBF83} with
experimental data in the polymer blend of polydimethylsiloxane and polyethylmethylsiloxane \cite{MMF92}.
As all these data are only available in a rather small range of $x$ we can apply the asymptotic theory
and avoid the discussion of the last paragraph. The use of a non-asymptotic theory
would also not be possible for a comparison with these experimental data for a second
reason: All data shown in Fig.~\ref{flossf12} were obtained for different temperatures
and wave vectors but these different values of $k$ and $\xi$ were not
indicated separately in the paper but only the corresponding value of $x=k\xi$.
This was also the reason why we could not fit the initial value of the Onsager coefficient
so that the only fit parameter, the prefactor of $\Omega(x)$, was set by the choice that
our result shall coincide with the result of Burstyn et al. in the limit of small values of $x$.
In addition we have to note that the experimental values for the function $\Omega(x)$ where
obtained from the data for the characteristic frequency dividing not by the full shear viscosity
depending on the correlation length but only by its constant background value so that we had
to correct this multiplying our theoretical expression for the function $\Omega(x)$ by $x^{-x_\eta}$.
In any case the experimental data shown in Fig.~\ref{flossf12} are not precise enough to favor
any of the presented theoretical expressions.

\section{The frequency dependent shear viscosity}

Since we have used information from the shear viscosity in the discussion of the light scattering line
width we shall add an analysis of the most recent shear viscosity data for Xe \cite{BMZ}  to
this paper. These new data allow a much more detailed analysis of the frequency dependent shear
viscosity leading to slightly different parameters than the discussion of the shear viscosity of
Xe in Ref.~\cite{fomo} which was based on older data. In Ref.~\cite{fomoflo} we have discussed the theoretical expression for the
frequency dependent shear viscosity, which is given by
\begin{equation}
\bar{\eta}(t,\Delta\rho,\omega)=\frac{k_BT}{4\pi}\frac{\xi_0}
{\ell f_t^2(\ell)\Gamma(\ell)}\left[1+E_t\big(f_t(\ell),v(\ell),w(\ell)\big)\right] \, , \label{eta2}
\end{equation}
with the one-loop perturbational contribution
\begin{eqnarray}\label{et}
&&E_t\big(f_t(\ell),v(\ell),w(\ell)\big)=-\frac{f_t^2}{96}
\Bigg\{1+6\Big[i\frac{v^2}{w}\ln v
+\frac{1}{v_+-v_-}\Big(\frac{v_-^2}{v_+}\ln v_--\frac{v_+^2}{v_-}
\ln v_+\Big)\Big]  \label{ET}\\
&-&\frac{4}{(v_+-v_-)^3}\Big[\frac{v_+^3-v_-^3}{3}+\frac{3}{2}(v_+-v_-)
(v_+^2\ln v_+ +v_-^2\ln v_-)-(v_+^3\ln v_+-v_-^3\ln v_-)\Big]
\nonumber \\
&+&\frac{2}{(v_+-v_-)^2}\Big[\frac{v_+^3}{v_-}(1+4\ln v_+)
+\frac{v_-^3}{v_+}(1+4\ln v_-) \nonumber \\
&+&\Big(\frac{1}{v_-}-\frac{2}{v_+-v_-}\Big)
\frac{v_+^4\ln v_+-v^4\ln v}{v_-}
+\Big(\frac{1}{v_+}+\frac{2}{v_+-v_-}\Big)
\frac{v_-^4\ln v_--v^4\ln v}{v_+}\Big]\Bigg\} \, .  \nonumber
\end{eqnarray}
The parameters introduced in Eq. (\ref{et}) are defined as

\begin{equation}\label{vplmi}
v(\ell)=\frac{\xi^{-2}(t)}{(\xi_0^{-1}\ell)^2} \ , \qquad
w(\ell,\omega)=\frac{\omega}{2\Gamma(\ell)(\xi_0^{-1}\ell)^4} \ ,
\end{equation}
\begin{equation}
v_\pm(\ell,\omega)=\frac{v}{2}\pm\sqrt{\left(\frac{v}{2}\right)^2+iw}\,,
\end{equation}
with the mode coupling $f_t(\ell)$ and the Onsager coefficient $\Gamma(\ell)$ given
by Eqs. (\ref{modecoupling}) and (\ref{onsager}). The mode coupling parameter $\ell$
is now a function of the correlation length $\xi$ and the frequency $\omega$ and results
from the solution of the matching condition
\begin{equation}\label{matchgeneral2}
\left(\frac{\xi_0}{\xi}\right)^8+\left(\frac{2\omega}{\Gamma(\ell)}
\right)^2=\ell^8 \, .
\end{equation}
At the moment of publication no experimental data were available to compare them
to our theoretical expressions. The situation has changed meanwhile as Berg et al. \cite{BMZ}
performed shear viscosity experiments at small frequencies in a microgravity environment onboard
a space shuttle. Comparing their experimental results with the mode coupling theory \cite{BF}
they found that they could only describe their data correctly multiplying the frequency by a factor of 2
in the theoretical expressions. They explained the introduction of this factor as a two-loop effect
correcting the errors of the one-loop expression used for the frequency dependent shear viscosity. With this
multiplicative factor for the frequency they were able to reproduce the experimental data
for the shear viscosity very well.

In Fig.~\ref{flossf07} and \ref{flossf08} we compare our theory with experimental data in microgravity
and in the earth's gravitational field \cite{BMZ,BM90} fitting the exponent $x_\eta$ with $f_0$ taken
from the light scattering experiments of Ref.~\cite{SH73}. In doing so we found $x_\eta=0.065$ instead of the value
$x_\eta=0.069$ used by Berg et al. We should note here that we can use the exponent $x_\eta=0.069$ (with
the initial values $f_0=0.959$ and $\Gamma_0=8.82\times10^{-18} {\rm cm^4/s}$) to get exactly the same theoretical
curves as shown in Fig.~\ref{flossf07} and \ref{flossf08} but then we are not able to describe the
characteristic frequency data correctly with this choice of $f_0$ and $\Gamma_0$. This fact that
the parameters $f_0$ and $x_\eta$ cannot be determined unambiguously from the shear viscosity
data alone was already discussed in detail in Ref.~\cite{fomoflo}.

In Fig.~\ref{flossf07} and \ref{flossf08} it turns out that we can describe the experimental data in microgravity
only if we multiply the frequency by a factor of 5, which may be justified for the same reason
as the factor of 2 in the mode coupling theory \cite{BMZ}. But then we are able to describe
not only the microgravity data but also the earth-bound experiments very well with a single
set of parameters shown in Tab~\ref{tab}. And once again let us mention that we have used the
same set of parameters to describe the characteristic frequency in Xe correctly in Fig.~\ref{flossf05} and \ref{flossf06}.
As the experimental data shown in Fig.~\ref{flossf08} cover a large range of reduced temperatures we had to add the
regular background contribution found in Ref.~\cite{BM90}, which is completely independent of the critical
behavior described within our model.

Berg et al. did not only measure the real part of the shear viscosity but determined also
the imaginary part of $\bar\eta$ from the phase shift. In Ref.~\cite{BMZ} they compared the
mode coupling result for the ratio $\Im(\bar\eta)/\Re(\bar\eta)$ with their experimental
data and found good agreement. Comparing our results with these experimental
data we get less satisfactory results \cite{lviv} because in our theory the ratio
$\Im(\bar\eta)/\Re(\bar\eta)$ approaches the finite value
\begin{equation}
\lim\limits_{T\to T_c} \frac{{\rm Im}(\bar\eta)}{{\rm Re}(\bar\eta)} =
\frac{1}{76}\,\frac{\pi}{2}\,
\left[1 - \frac{1}{76}\left\{3\ln(1/4)-1/3\right\}\right]^{-1} \approx 0.0195 \label{limit}
\end{equation}
at $T_c$ which is different from the value $0.0353$ obtained from the mode coupling theory with
the exponent $x_\eta=0.069$ \cite{BMZ} which turns out to be in good agreement with the experimental
data. As the limit of the ratio $\Im(\bar\eta)/\Re(\bar\eta)$ does not contain any free parameter at
$T_c$ it cannot be improved and the deviation of our theory from the experiments may be explained
by the fact that a one-loop order perturbation theory is not able to describe such small effects
(the imaginary part of the shear viscosity is only about 3\%
of the total complex shear viscosity) and therefore a two-loop theory may be expected to yield
much better agreement. In this respect we should also note that the mode coupling expression used
by Berg et al. is not purely of one-loop order since it makes use of the experimental value for the exponent
$x_\eta$ which differs significantly from its one-loop value. If we insert the one-loop value $x_\eta=1/19$
into the mode coupling expressions we would get a limit $\Im(\bar\eta)/\Re(\bar\eta)\approx0.0271$ at $T_c$ which
is also significantly lower than the measured limiting ratio. So the main difference between
the mode coupling theory and our theory is, that it is not possible to introduce the true critical exponent $x_\eta$
in our expression for $\Im(\bar\eta)/\Re(\bar\eta)$ and therefore deviations from the one-loop order
perturbation theory cannot be weakened by the use of the correct value for $x_\eta$.

\section{Conclusion}

We were able to show that our one-loop perturbation theory result for the characteristic frequency evaluated
within the field theoretical method of the renormalization group theory does not only reproduce the
correct wave vector and correlation length dependence in the hydrodynamic region as well as in the
critical region but is also able to describe experimental data correctly for a large range of wave vectors
and reduced temperatures. In addition we showed that also the result for the shear viscosity evaluated
within the same model is in good agreement with experiments if a two-loop value for the critical exponent
is taken.

There are however some points which indicate the need for a two-loop analysis of the model: First we have
seen that in one-loop order the dynamic correlation function is always of Lorentzian form whereas scaling
theory \cite{sofia} predicts deviations for large frequencies. Second we are not able to get the experimental
limiting value for the ratio of the imaginary and real part of the frequency dependent shear viscosity
$\Im(\bar\eta)/\Re(\bar\eta)$ at $T_c$ and we have to introduce a multiplicative  factor for the frequency
in order to describe the experimental data correctly. This makes a profound two-loop
analysis inevitable which is currently in progress.

{\bf Acknowledgment:} This  work was supported by the
Fonds zur F\"orderung der wissenschaftlichen Forschung under Project No. P12422-TPH.

\begin{center}
{\bf Figure caption}
\end{center}

\begin{figure}[ht]
     \centering
       \vspace{-1cm}
       \epsfig{file=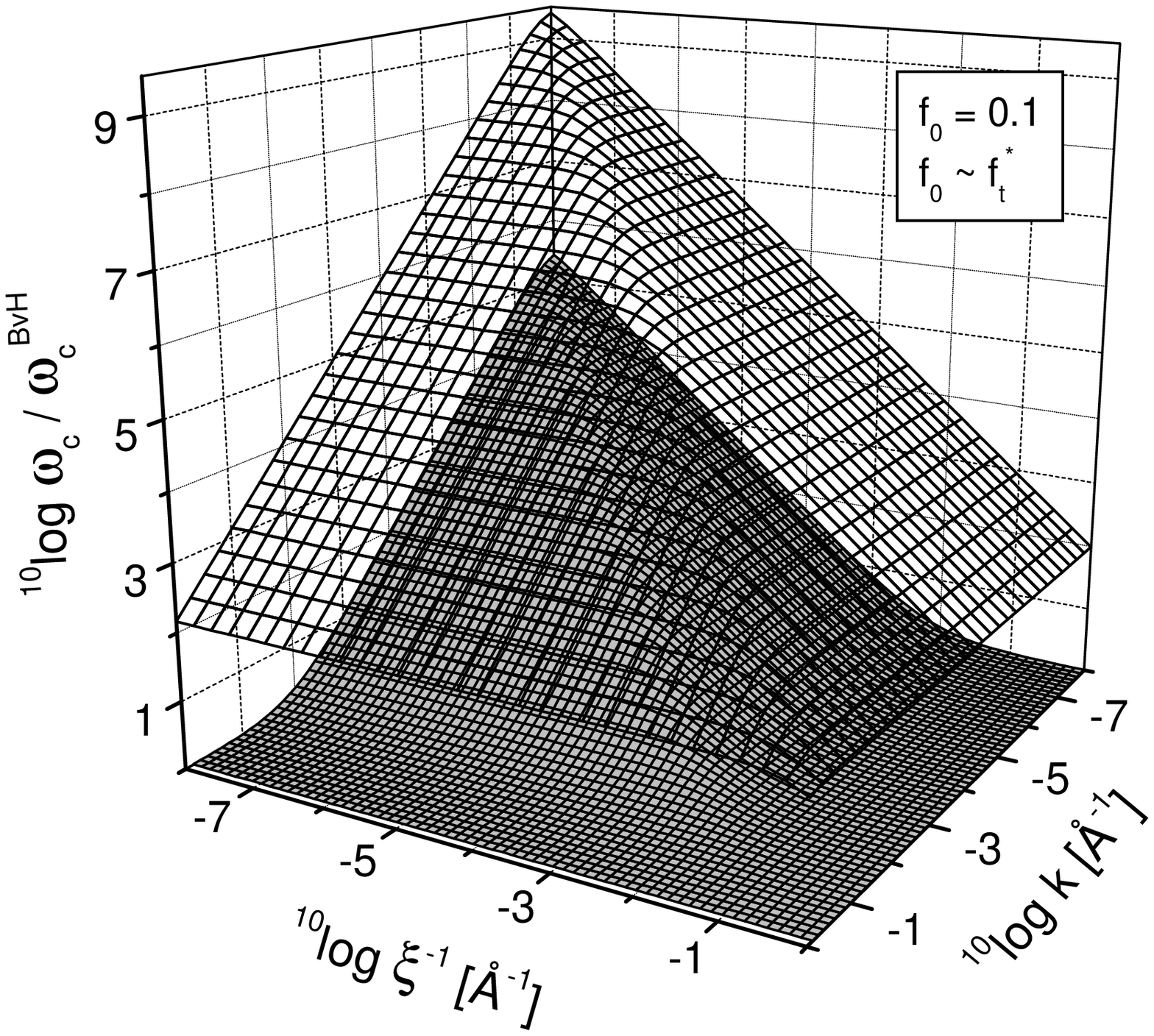,height=10cm}
       \vspace{-1cm}
    \caption{ \label{flossf3a}}
  \end{figure}

{\bf Fig. \ref{flossf3a}}

Ratio of the characteristic frequency $\omega_c$ divided by the van Hove background expression $\omega_c^{BvH}$ for
$f_0=0.1$ or $k_0=7.98\times10^{-3}\AA^{-1}$ respectively where the ratio becomes 1 in the background limit $\xi\to0$
and $k\to\infty$ and for $f_0\approx f_t^*$ where the van Hove expression is never reached by the asymptotic
characteristic frequency.

\begin{figure}[ht]
     \centering
       \epsfig{file=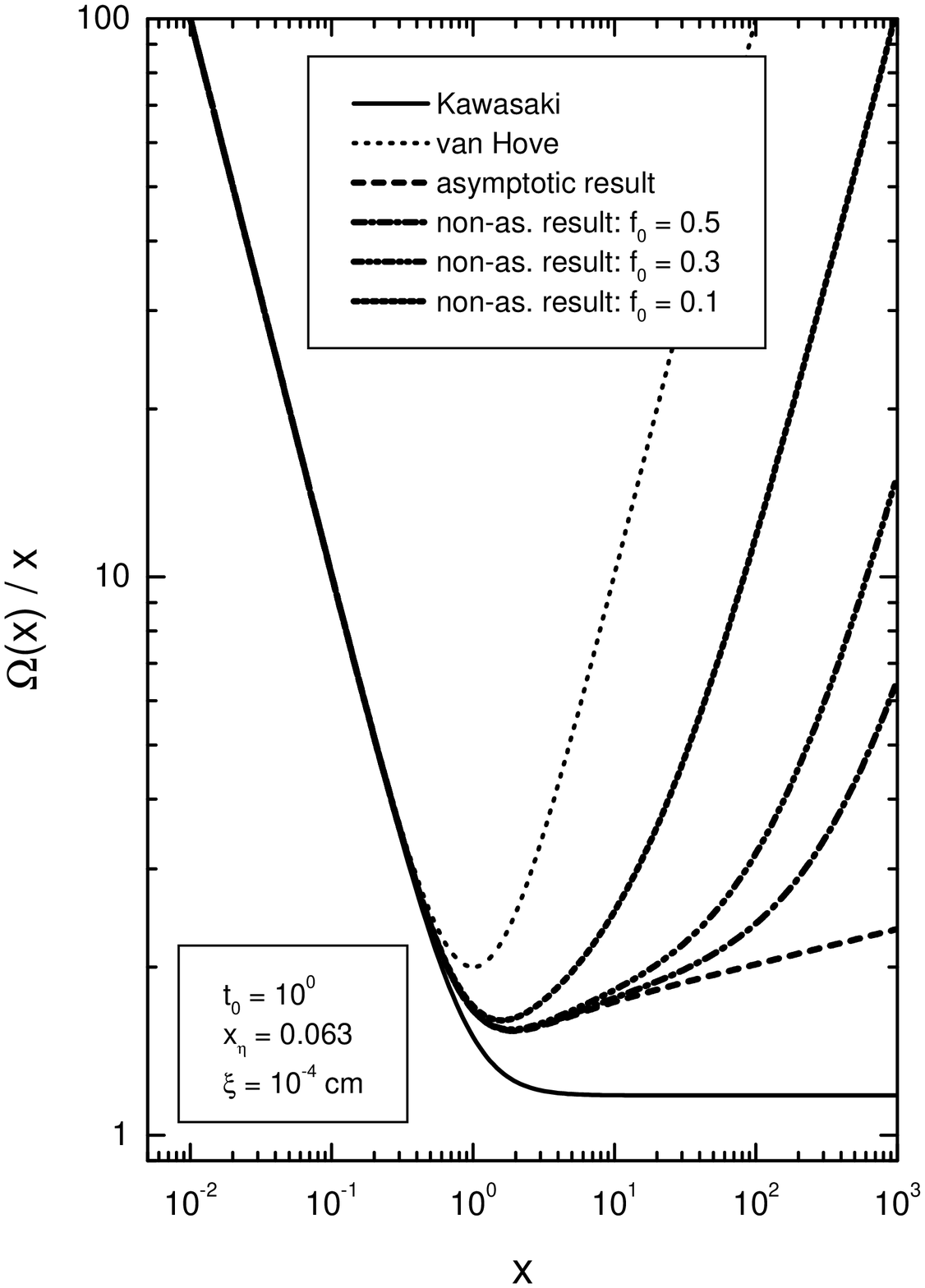,width=7cm,height=8cm}
     \caption{ \label{flossf03}}
  \end{figure}

{\bf Fig. \ref{flossf03}}

Comparison of our asymptotic and non-asymptotic (for various values of $f_0$ at constant
correlation length $\xi$) results for $\Omega(x)/x$ with the Ornstein-Zernike theory and
the theoretical result of Kawasaki \cite{KL72}.

\begin{figure}[ht]
     \centering
       \epsfig{file=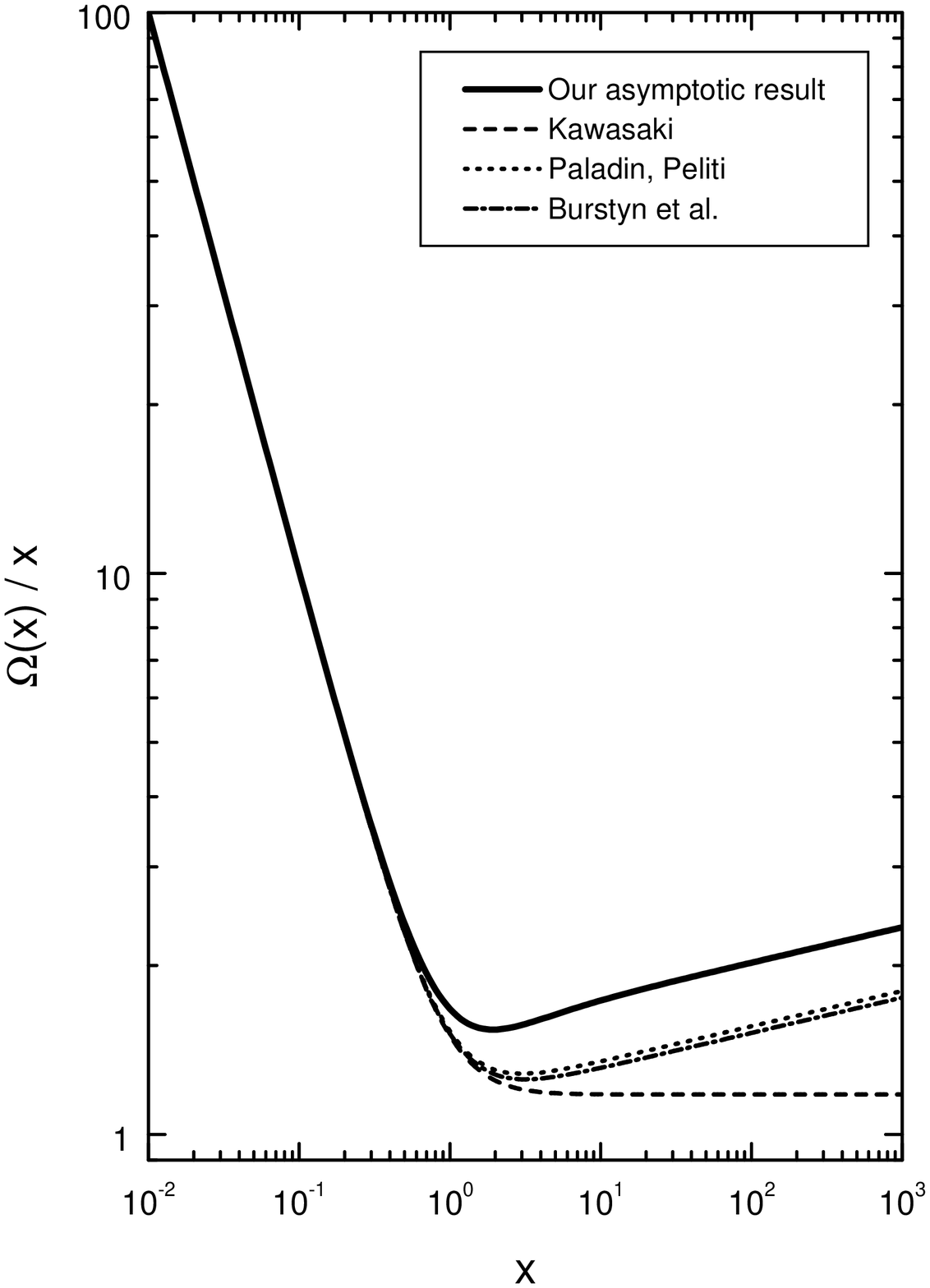,width=7cm,height=8cm}
     \caption{ \label{flossf04}}
  \end{figure}

{\bf Fig. \ref{flossf04}}

Comparison of our asymptotic result for $\Omega(x)/x$ with the theoretical
results of Kawasaki \cite{KL72}, Paladin and Peliti \cite{PP82}, and Burstyn et al. \cite{BSBF83}.

\begin{figure}[ht]
     \centering
       \vspace{-1cm}
       \epsfig{file=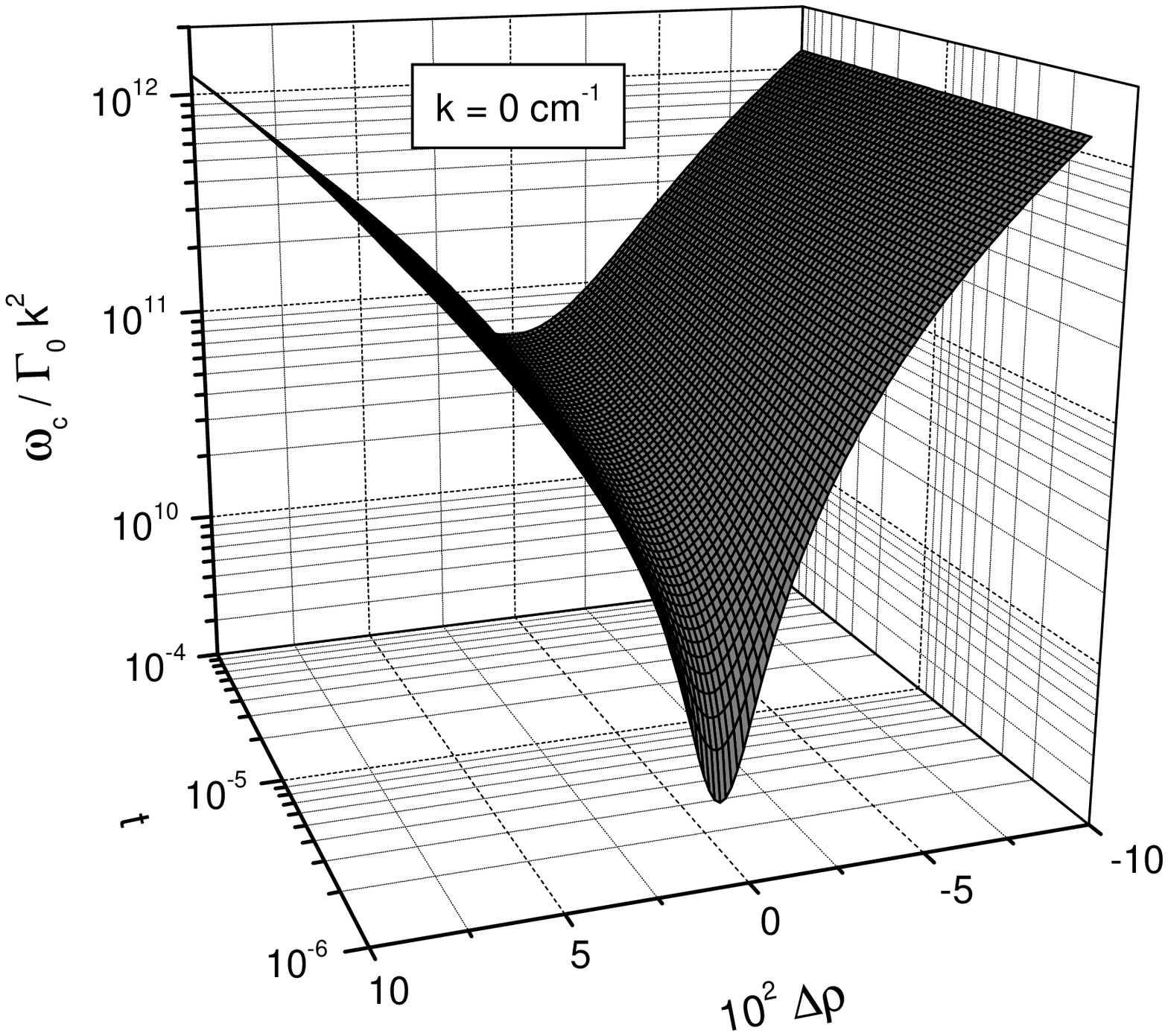,height=10cm}
       \vspace{-1cm}
    \caption{ \label{flossf02}}
  \end{figure}

{\bf Fig. \ref{flossf02}}

The characteristic frequency $\omega_c/\Gamma_0k^2$ as a function of the reduced temperature $t$ and the
reduced density $\Delta\rho$ in the hydrodynamic limit for $k=0$.

\begin{figure}[ht]
     \centering
       \epsfig{file=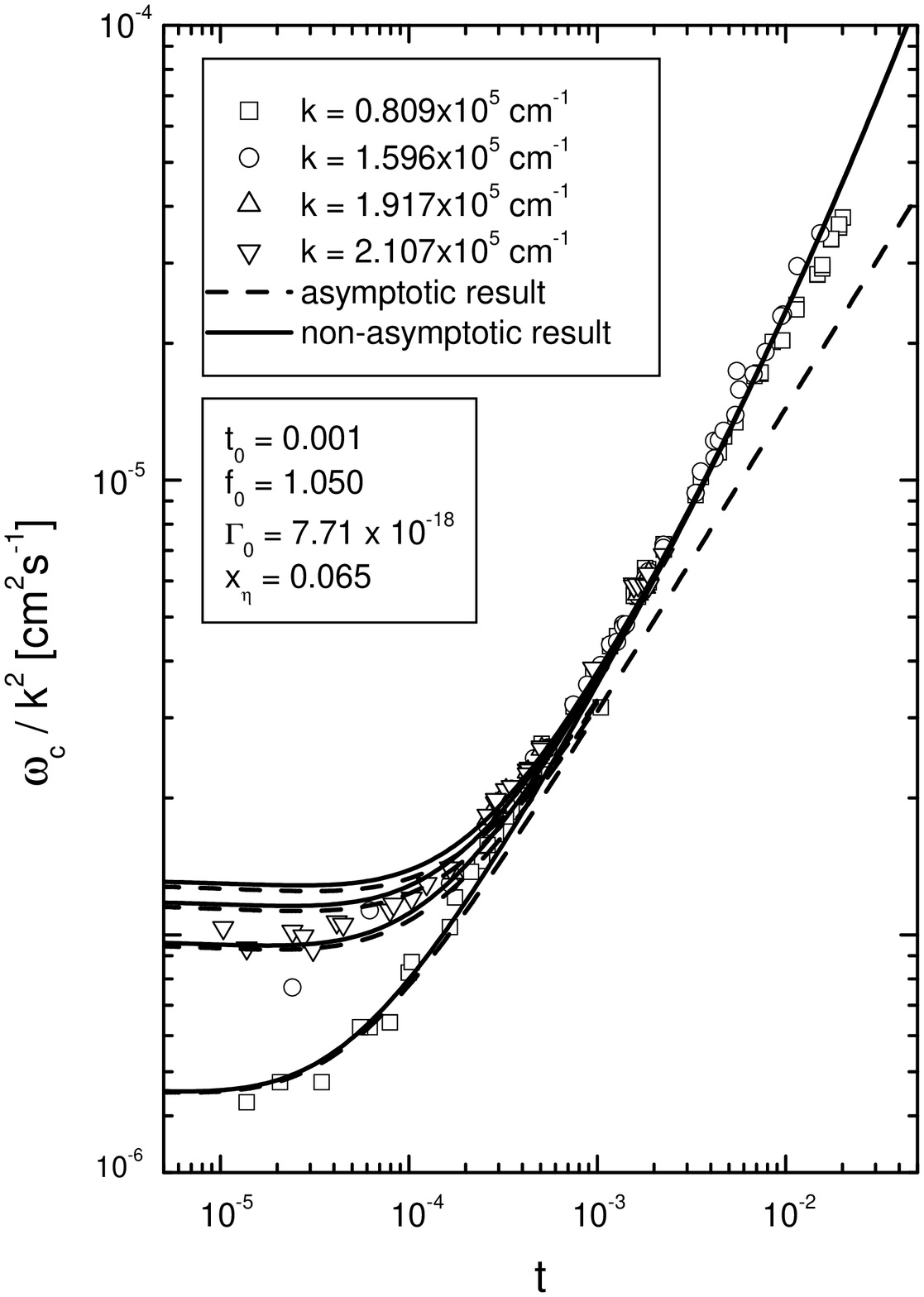,width=7cm,height=8cm}
     \caption{ \label{flossf05}}
  \end{figure}

{\bf Fig. \ref{flossf05}}

Comparison of the asymptotic (full lines) and non-asymptotic (dashed lines) expressions for $\omega_c/k^2$
with the ${\rm Xe}$-data of Ref.~\cite{SH73}.

\begin{figure}[ht]
     \centering
       \epsfig{file=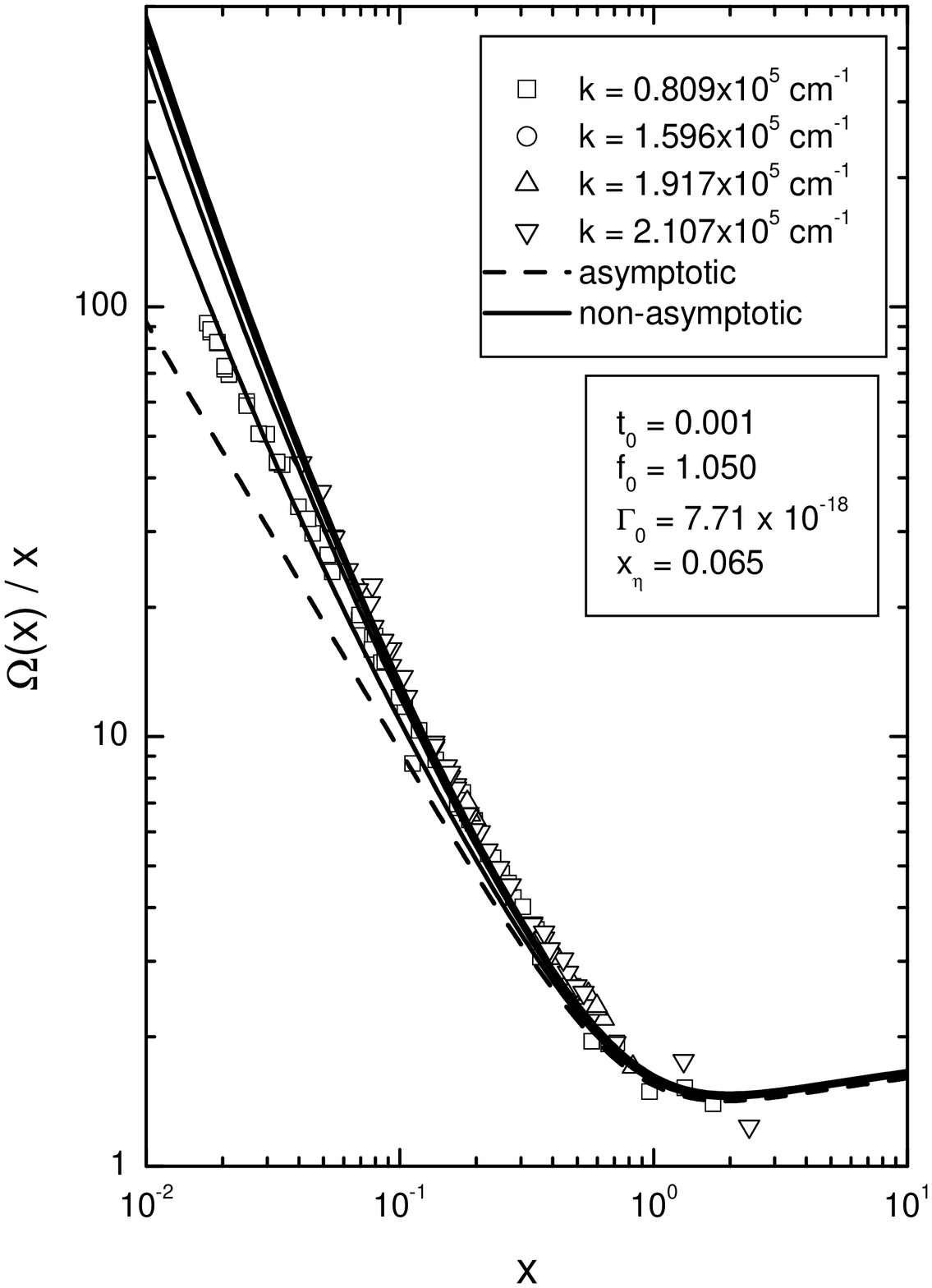,width=7cm,height=8cm}
     \caption{ \label{flossf06}}
  \end{figure}

{\bf Fig. \ref{flossf06}}

Comparison of the asymptotic (full lines) and non-asymptotic (dashed lines) expressions for $\Omega(x)/x$
with the ${\rm Xe}$-data of Ref.~\cite{SH73}.

\begin{figure}[ht]
     \centering
       \epsfig{file=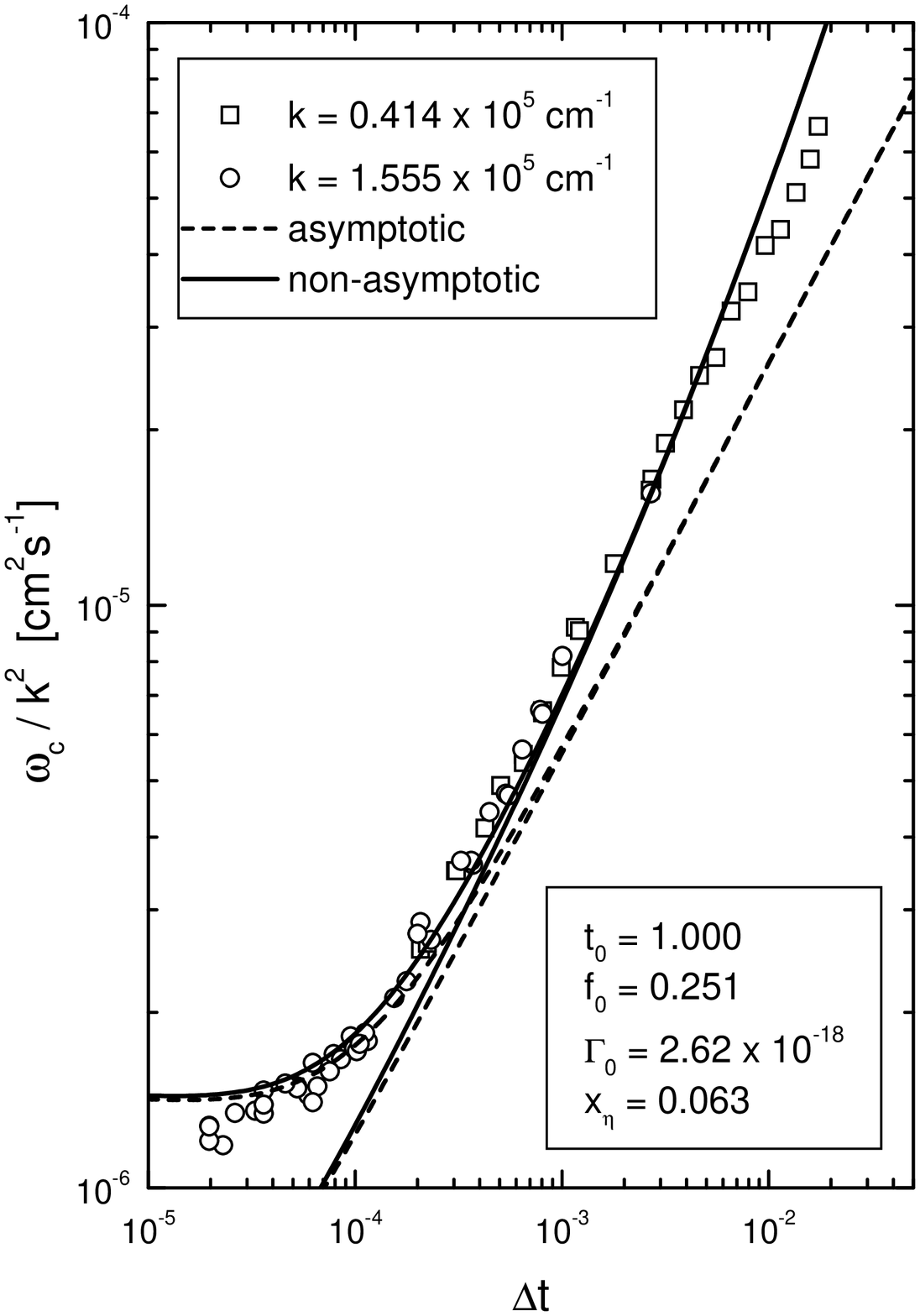,width=7cm,height=8cm}
     \caption{ \label{flossf09}}
  \end{figure}

{\bf Fig. \ref{flossf09}}

Comparison of the asymptotic (full lines) and non-asymptotic (dashed lines) expression for $\omega_c/k^2$
with the ${\rm CO_2}$-data of Ref.~\cite{SH73}.

\begin{figure}[ht]
     \centering
       \epsfig{file=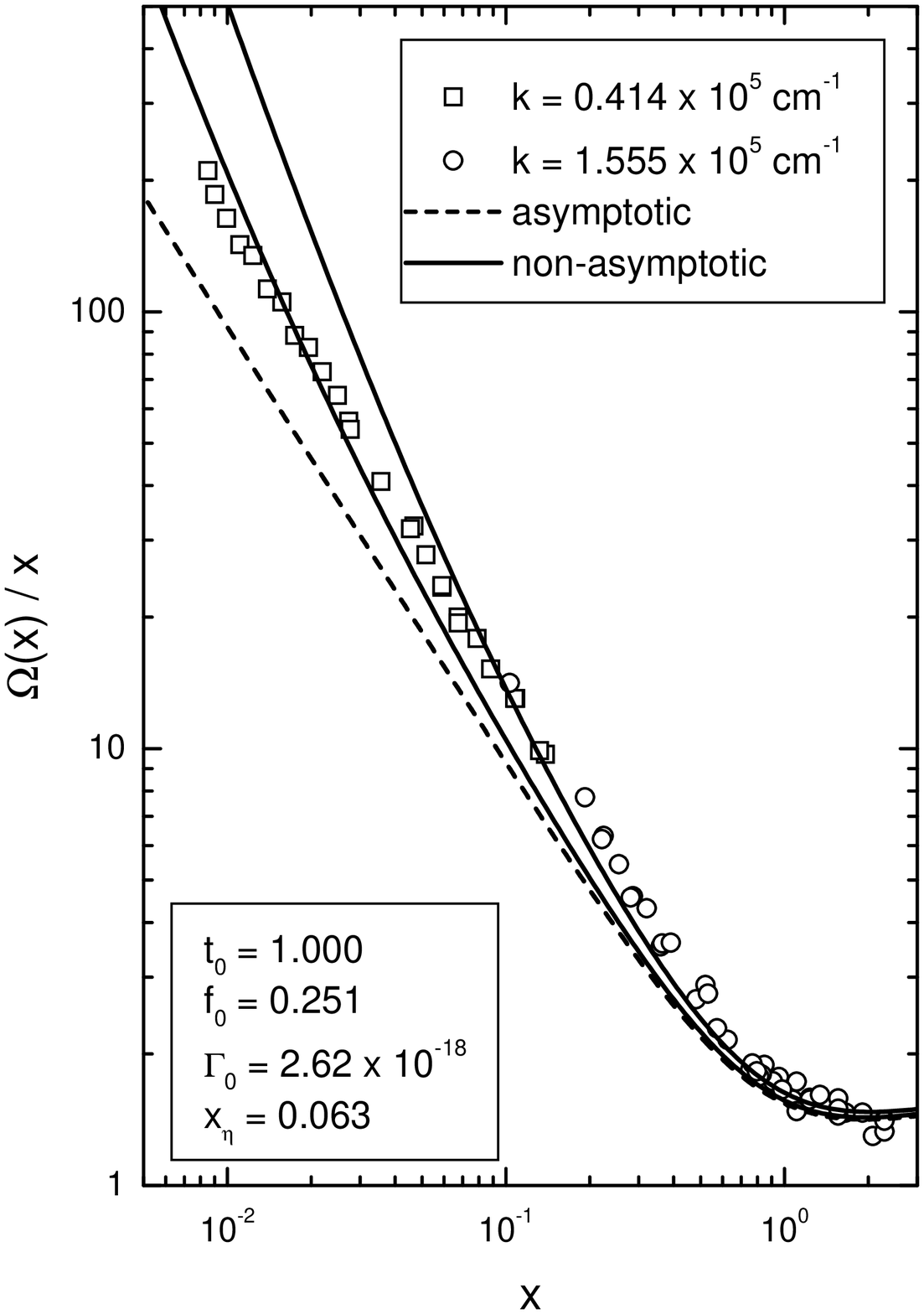,width=7cm,height=8cm}
     \caption{ \label{flossf10}}
  \end{figure}

{\bf Fig. \ref{flossf10}}

Comparison of the asymptotic (full lines) and non-asymptotic (dashed line) expressions for $\Omega(x)/x$
with the ${\rm CO_2}$-data of Ref.~\cite{SH73}.

\begin{figure}[ht]
     \centering
       \epsfig{file=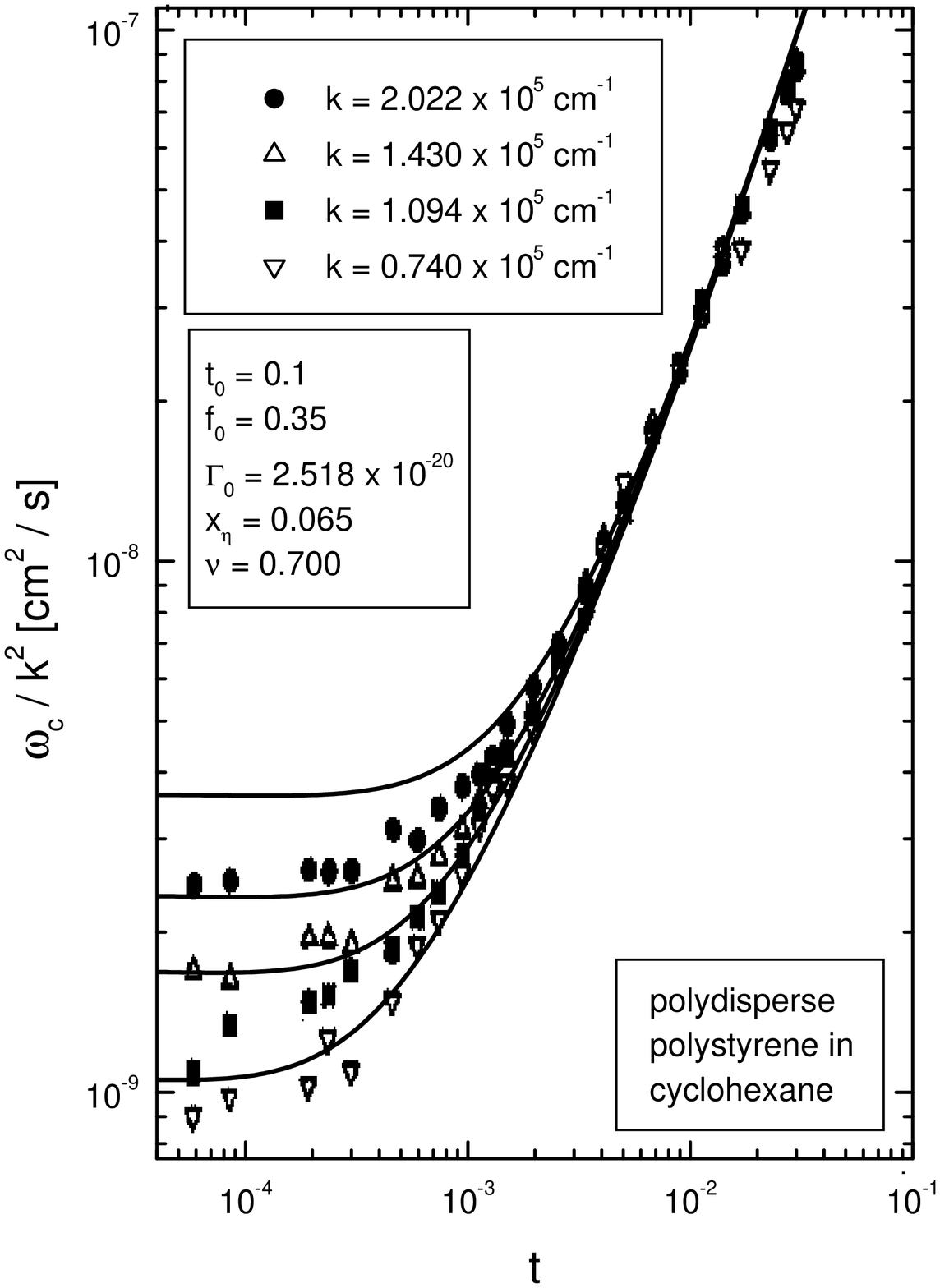,width=7cm,height=8cm}
     \caption{ \label{flossf11}}
  \end{figure}

{\bf Fig. \ref{flossf11}}

Comparison of the non-asymptotic characteristic frequency $\omega_c/k^2$ with the experimental data
of Ref.~\cite{KKD98} in a polymer solution after subtracting the regular background.

\begin{figure}[ht]
     \centering
       \epsfig{file=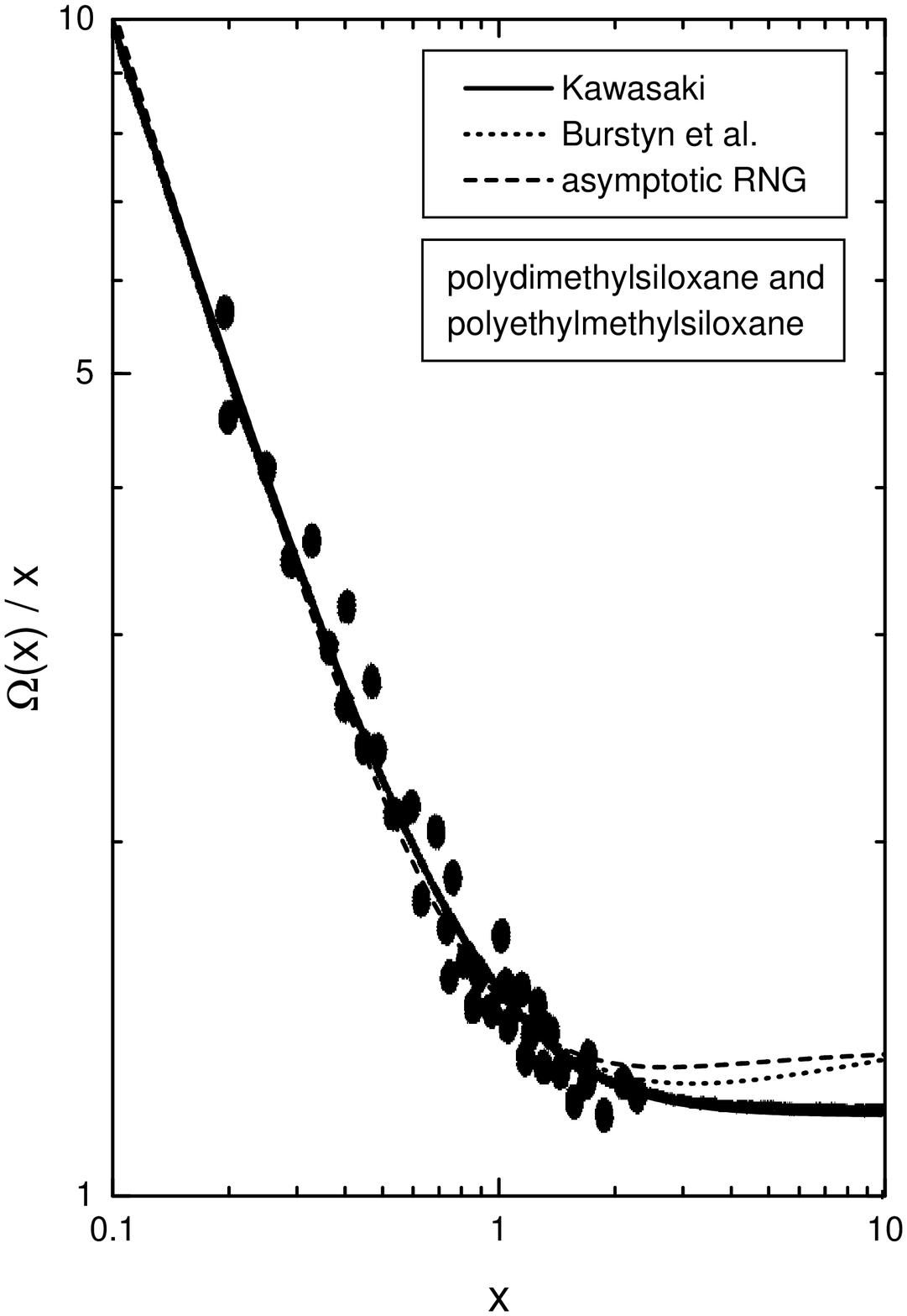,width=7cm,height=8cm}
     \caption{ \label{flossf12}}
  \end{figure}

{\bf Fig. \ref{flossf12}}

Comparison of the asymptotic expression for $\Omega(x)/x$ with the experimental data of Ref.~\cite{MMF92}  
in a polymer mixture.

\begin{figure}[ht]
     \centering
       \epsfig{file=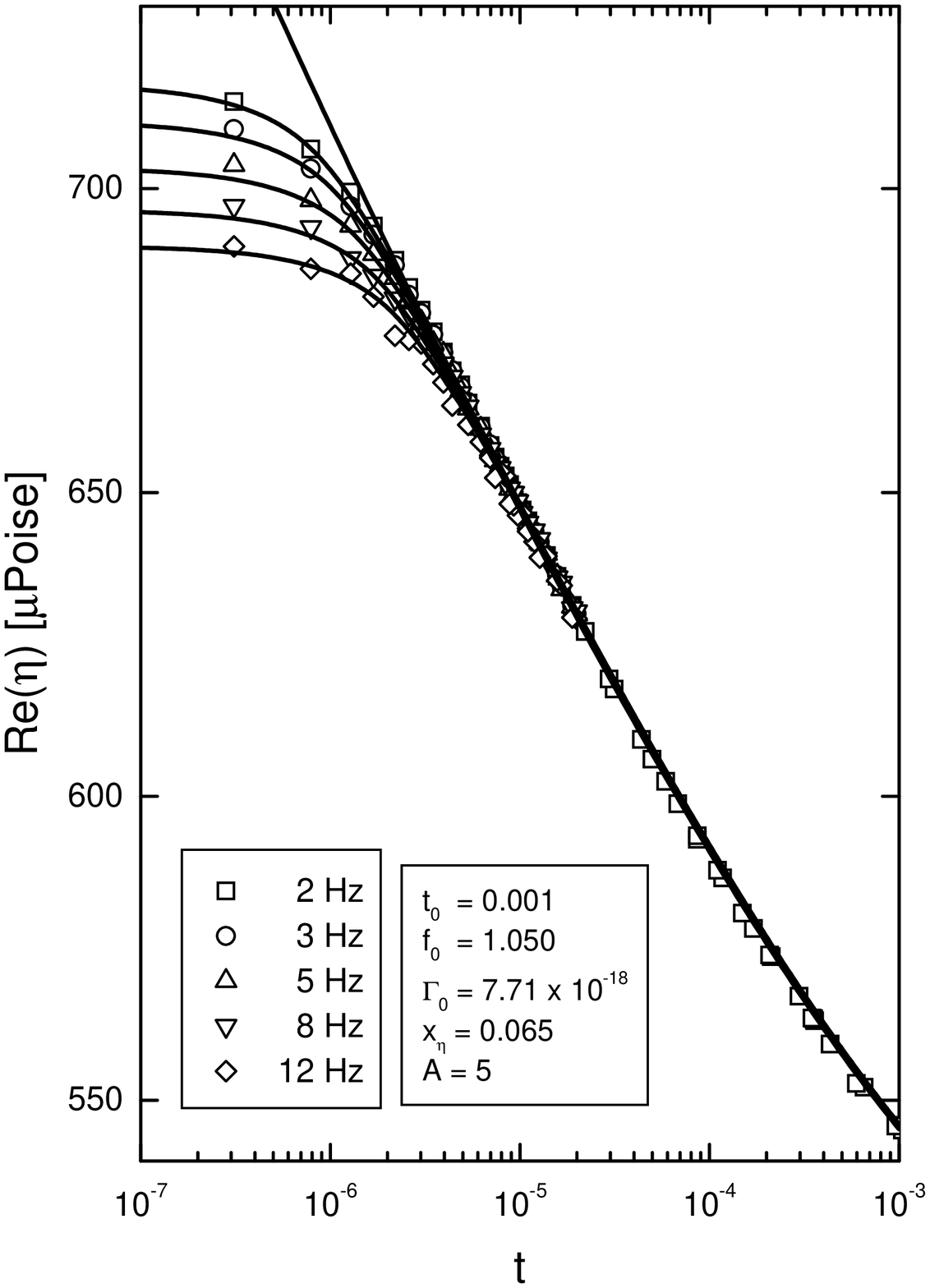,width=7cm,height=8cm}
     \caption{ \label{flossf07}}
  \end{figure}

{\bf Fig. \ref{flossf07}}

Comparison of the theoretical expression for the real part of the shear viscosity in microgravity
at various frequencies with the experimental data of Ref.~\cite{BMZ}. See text for details.

\begin{figure}[ht]
     \centering
       \epsfig{file=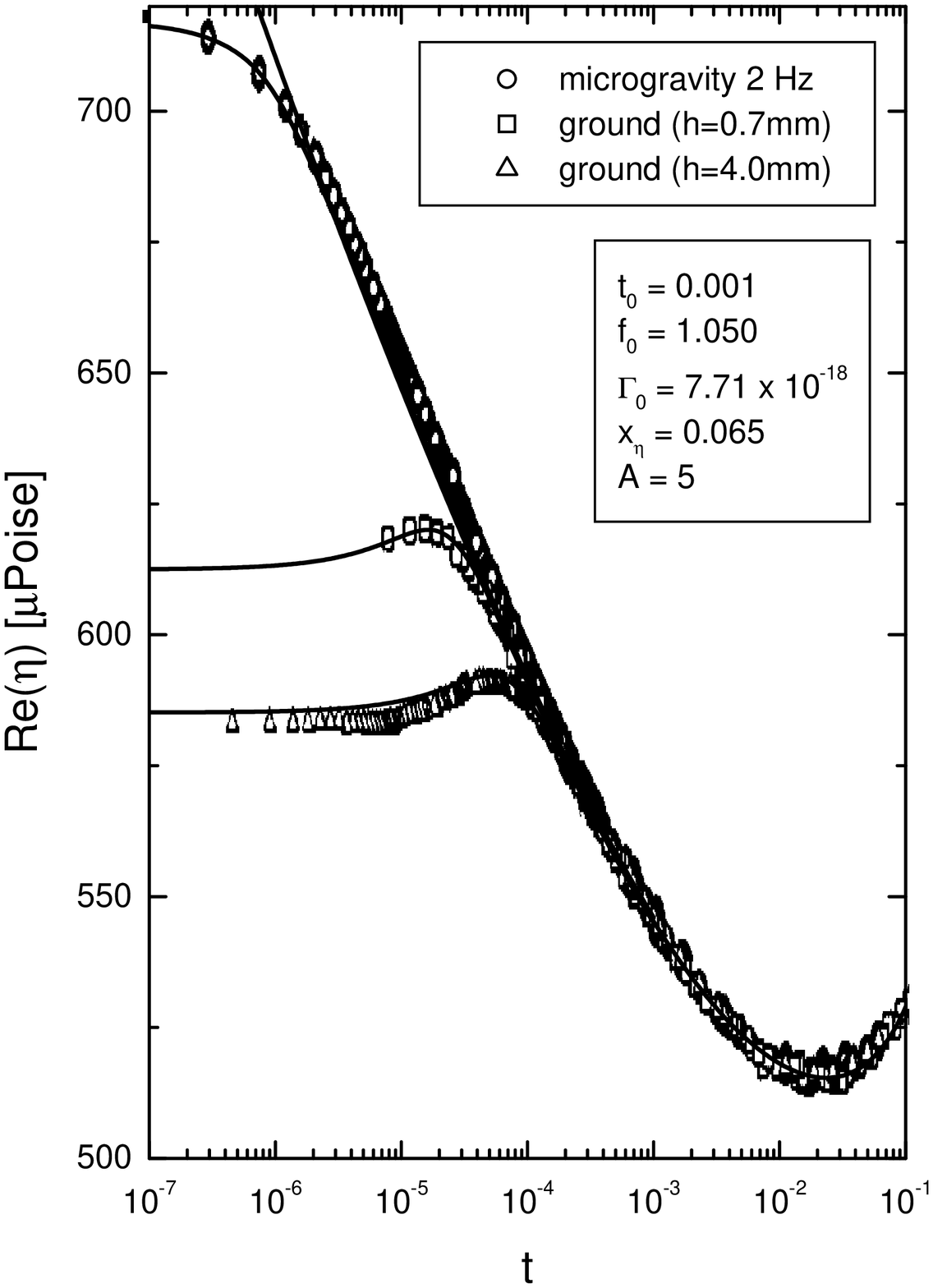,width=7cm,height=8cm}
     \caption{ \label{flossf08}}
  \end{figure}

{\bf Fig. \ref{flossf08}}

Comparison of the theoretical expressions for the real part of the shear viscosity in microgravity (at frequencies 0Hz and 2Hz)
as well as in earthbound experiments (for two different vessel heights) with the experimental data of Ref.~\cite{BMZ,BM90}.

\begin{center}
{\bf Table caption}
\end{center}

{\bf Table\ref{tab}}

Nonuniversal parameters of Xe.


\narrowtext
\begin{table}
\begin{center}
\begin{tabular}{cccccccc}
Liquid & $\xi_0 [{\AA}]$ & $t_0$ & $f_0$ & $\Gamma_0 [cm^4/s]$ & $x_\eta$ & $\nu$ & $k_0 [{\rm cm}^{-1}]$ \\ \hline
${\rm Xe}$   & 1.84 & 0.001 & 1.050 & $7.71\times10^{-18}$ & 0.065 & 0.63 & $48.0\times10^5$  \\
${\rm CO_2}$ & 1.60 & 1.000 & 0.251 & $2.62\times10^{-18}$ & 0.063 & 0.63 & $32.8\times10^5$  \\
PDPS         & 4.60 & 0.100 & 0.350 & $2.52\times10^{-20}$ & 0.065 & 0.70 & $4.66\times10^5$
\end{tabular}
\end{center}
\caption{\label{tab} }
\end{table}

\end{document}